\documentclass[12pt]{article}
\usepackage[dvips]{graphicx}
\usepackage{amssymb}
\usepackage{amsmath}
\usepackage{epsfig}
\usepackage{cite}

\numberwithin{equation}{section}
\numberwithin{table}{section}

\def\beq{\begin{equation}}
\def\eeq{\end{equation}}
\def\be{\begin{equation}}
\def\ee{\end{equation}}
\def\bea{\begin{eqnarray}}
\def\eea{\end{eqnarray}}

\def\cM{\mathcal{M}}

\def\Re{{\rm Re\,}}
\def\Im{{\rm Im\,}}

\def\f{{\mathfrak{f}}}
\def\vev#1{{\langle #1\rangle}}
\def\rep#1{{\bf #1}}
\def\SO{\textrm{SO}}
\def\SU{\textrm{SU}}
\DeclareRobustCommand{\SkipTocEntry}[4]{}
\RequirePackage{color}

\def\vev#1{{\langle #1\rangle}}

\newcommand\iu{\operatorname{i}}

\begin{document}
\begin{titlepage}
\begin{center}
\rightline{\small ZMP-HH/14-15}
\rightline{\small CERN-PH-TH/2014-103}

\vskip 1cm

{\Large \bf
 Maximally Supersymmetric $\bf{AdS_4}$ Vacua in $N=4$ Supergravity}
\vskip 1.2cm

{\bf  Jan Louis$^{a,b}$ and Hagen Triendl$^{c}$ }

\vskip 0.8cm

$^{a}${\em Fachbereich Physik der Universit\"at Hamburg, Luruper Chaussee 149, 22761 Hamburg, Germany}
\vskip 0.3cm

{}$^{b}${\em Zentrum f\"ur Mathematische Physik,
Universit\"at Hamburg,\\
Bundesstrasse 55, D-20146 Hamburg, Germany}
\vskip 0.3cm

$^e${\em Theory Division, Physics Department, CERN, CH-1211 Geneva 23, Switzerland}

\vskip 0.3cm

{\tt jan.louis@desy.de, hagen.triendl@cern.ch}

\end{center}

\vskip 1cm

\begin{center} {\bf ABSTRACT } \end{center}

\noindent

We study AdS backgrounds of
$N=4$ supergravity in four space-time dimensions which
preserve all sixteen supercharges.
We show that the graviphotons have to
form a subgroup of the gauge group that consists of an electric and a magnetic
$\SO(3)_+\times \SO(3)_-$. Moreover, these $N=4$ AdS backgrounds are necessarily isolated points in field space
which have no moduli.

\vfill


June 2014

\end{titlepage}


\tableofcontents


\section{Introduction}

The maximally symmetric space-time backgrounds of
supergravity theories which preserve some of the supercharges are
either anti--de Sitter (AdS) or Minkowski (M) spaces.
It is of interest to
study such backgrounds and find the (model-independent)
properties of the associated moduli spaces.
In $N=2$ supergravities in four space-time dimensions ($d=4$)
the fully supersymmetric AdS$_4$ backgrounds were determined
in \cite{Hristov:2009uj,Louis:2012ux} while the structure of the moduli space
of $N=1$ and $N=2$ AdS$_4$ backgrounds was given in
\cite{deAlwis:2013jaa}.
It was found that generically  a supersymmetric  AdS$_4$ background
of $N=1$ and $N=2$ supergravity has no moduli space. However,
by appropriately tuning the mass parameters of the theory flat directions
which preserve all supercharges
may occur. In $N=1$ they span a field space
which  is necessarily real and has at best
half the dimension of the original field space.
In $N=2$  the moduli space is a K\"ahler manifold --
again at best of half the dimension of the original field space.
Both results are in agreement
with the AdS/CFT correspondence which relates
these  backgrounds to superconformal field theories on the $d=3$
boundary of  AdS$_4$ with multiplets which are in representations
of theories that have only half of the supercharges.\footnote{The other half
are the superconformal supercharges.}

For $N=4$ supergravity in $d=4$ an analogous investigation is lacking so far
and it is the purpose of this paper to close this gap.
In contrast to gauged supergravities with eight or less supercharges,
in $N=4$ supergravity the mass parameters cannot be freely tuned
and are determined by the choice of the gauge group. This in turn
suggests that the
dimension and structure of the moduli space is also fixed.
From the AdS/CFT perspective one expects the moduli space
to be hyper-K\"ahler -- if it
exists at all.\footnote{It has been conjectured \cite{ofer}
that  $d=3$ superconformal field theories with eight supercharges
(plus eight conformal supercharges) have no exactly marginal deformations
which would in turn suggest that there is
no moduli space in an $N=4$ AdS$_4$ bulk
supergravity.
We thank O.\ Aharony for this inspirational remark which
prompted the present investigation.}

After the initial construction of (electrically) gauged $N=4$ supergravity
\cite{Chamseddine:1980cp,deRoo:1984gd,Bergshoeff:1985ms,deRoo:1985np,deRoo:1985jh} it was shown
that within this class of theories no supersymmetric AdS-backgrounds exist
and any background preserving some supercharges has to be Minkowskian M$_4$
\cite{deRoo:1986yw,Wagemans:1987zy}.\footnote{For a recent analysis of partial $N=4$ breaking see, for example, \cite{Horst:2012ub} and references therein.}
However the same papers realized that
supersymmetric AdS-backgrounds can occur when additional parameters
 are non-trivial.  These de~Roo-Wagemans angles
gauge isometries  with respect to dual magnetic vector multiplets.
Generic gauged supergravities including magnetic vector multiplets
have been constructed
in \cite{deWit:2005ub} introducing what is now called the
embedding tensor formalism. This was subsequently used in
\cite{Schon:2006kz} to construct the most general gauged
$N=4$ supergravity coupled to vector multiplets in $d=4$.
It is within this framework that we conduct our analysis.

We find that the existence of a fully supersymmetric AdS$_4$ background
imposes a set of constraints on the embedding tensor.
They in turn imply that the complex scalar $\tau$ of the gravitational multiplet
has to be uncharged and they also restrict the possible gauge groups~$G_0$.
More precisely,
the six graviphotons of $N=4$ supergravity have to gauge
an unbroken $\textrm{SO}(3)_+ \times\textrm{SO}(3)_-$ inside the R-symmetry
$\SO(6)_R$ where one of the factors is electric while the other is magnetic.
In general the two factors can be part of a larger
gauge group with the structure
$G_0 = G_+\times G_-\times G_0^v \subset \textrm{SO}(6,n)$
where $G_0^v \subset SO(n)$ is a separate factor. In the $N=4$ AdS$_4$ vacuum the group $G_+\times G_-$ is spontaneously
broken to its maximal compact subgroup, containing the two
$\textrm{SO}(3)_\pm$ factors. Furthermore, the potential has supersymmetric flat directions which, however, are precisely the Goldstone bosons
of the spontaneous symmetry breaking. No further flat directions
and thus no moduli space exists.

This paper is organized as follows.
In Section~\ref{Prelim} we recall the properties of $N=4$ gauged supergravity
that we need for our analysis.
In Section~\ref{AdS} we analyze $N=4$ AdS$_4$ backgrounds
and determine the constraints on the embedding tensor.
We then show that an
$\textrm{SO}(3)_+\times \textrm{SO}(3)_-$ subgroup of the R-symmetry
group  is necessarily gauged and we also determine the allowed structure
of the full gauge group $G_0$.
In Section~\ref{moduli} we show that the conditions
for an $N=4$ AdS-background admit a set of flat directions
corresponding to the  Goldstone bosons of the spontaneously broken
$G_0$. However, no further flat directions do exist
which indeed confirms that the backgrounds found in
Section~\ref{AdS} are isolated points in the scalar field space.
Some of the technical analysis is relegated to
three appendices.

\section{Preliminaries: $N=4$ gauged supergravity}\label{Prelim}

Let us set the stage and recall the properties of $d=4, N=4$
gauged supergravity \cite{Schon:2006kz} which are relevant
in the following. A generic $N=4$
spectrum consists of the gravity multiplet together with $n$~vector
multiplets.  The gravity multiplet contains the graviton $g_{\mu\nu}$, four
gravitini $\psi^i_{\mu},\, i=1,\ldots,4$, six vectors $A_{\mu}^m,\,
m=1,\ldots,6$, four spin-1/2 fermions $\chi^i$ and the complex scalar~$\tau$.
We label the vector multiplets with the index $a=1,\ldots,n$ and each multiplet
contains a vector $A_{\mu}^a$, four spin-1/2 gauginos $\lambda^{ai}$
and $6$ scalars $\phi^{am}$.
So altogether the spectrum features the graviton,
4 gravitini, $(6+n)$ vector bosons,
$(4+4n)$~spin-1/2~fermions and $(6n+2)$ scalars.

The field space $\cM$ of the scalars is the coset
\be\label{N4coset}
\cM = \frac{SL(2)}{\SO(2)}\times  \frac{\SO(6,n)}{\SO(6)\times \SO(n)}\ ,
\ee
where the first factor is spanned by $\tau$ while the second factor
is spanned by the scalars $\phi^{am}$ in the vector multiplets.
Both cosets are conveniently parametrized by vielbein fields.
For the first factor the vielbein is the complex vector
$\nu_\alpha$, $\alpha=+,-$,
which reads in terms of $\tau$ as
\begin{equation}\label{nutau}
\nu_\alpha\ =\ \frac{1}{\sqrt{\Im \tau}}\, \left( \begin{aligned} \tau \\ 1 \end{aligned} \right) \ ,
\end{equation}
and defines 
\be\label{MSL2}
M_{\alpha \beta} = \Re ( \nu_\alpha (\nu_\beta)^*)\ ,\qquad
\epsilon_{\alpha \beta} = \Im ( \nu_\alpha (\nu_\beta)^*)\ .
\ee
The second factor in \eqref{N4coset} is parametrized by
the vielbein $\nu =(\nu^m_M, \nu^a_M), M=1,\ldots,n+6$
which is an element of $\SO(6,n)$ and thus obeys
\begin{equation} \label{eq:vielbein_metric}
 \eta_{MN} = - \nu^m_M \nu^m_N + \nu^a_M \nu^a_N \ ,
\end{equation}
where  $\eta_{MN}={\rm diag}(-1,-1,-1,-1,-1,-1,+1,\dots,+1)$
is the flat $\SO(6,n)$ metric.
The metric on the coset is then given by
\begin{equation}
M_{MN}\ =\ \nu^m_M \nu^m_N + \nu^a_M \nu^a_N \ =\ 2 \nu^m_M\nu^m_N + \eta_{MN} \ .
\end{equation}

The couplings of $N=4$ gauged supergravity depend on two field-independent
$SL(2)\times \SO(6,n)$-tensors (called embedding tensors) denoted by
$\xi_{\alpha M}$ and $f_{\alpha [MNP]}$. Their entries are real numbers and
supersymmetry imposes
a set of coupled consistency conditions on both tensors
known as the quadratic constraints \cite{Schon:2006kz}
\begin{equation}\label{eq:quadconstr}\begin{aligned}
 \xi^M_\alpha \xi_{\beta M} =  0 \ ,& \qquad
\epsilon^{\alpha \beta} (\xi^P_\alpha f_{\beta PMN} + \xi_{\alpha M} \xi_{\beta N}) =  0 \ ,
 \\
 \xi^P_{(\alpha} f_{\beta) PMN} =  0 \ , &\qquad
 3 f_{\alpha R[MN} f_{|\beta| PQ]}{}^R + 2 \xi_{(\alpha[M} f_{\beta) NPQ]} =  0 \ ,
 \\
 \epsilon^{\alpha \beta} (f_{\alpha MNR} f_{\beta PQ}{}^R& - \xi^R_\alpha f_{\beta R[M[P} \eta_{Q]N]} - \xi_{\alpha[M} f_{|\beta|N]PQ} + \xi_{\alpha[P} f_{|\beta|Q]MN} ) =  0 \ .
\end{aligned} \end{equation}
Their solutions parametrize the different consistent
$N=4$ theories and in particular determine the gauge group,
the order parameters for spontaneous supersymmetry breaking and the potential.

The full bosonic Lagrangian is recorded in  \cite{Schon:2006kz} but
for the analysis in this paper we only need the  potential $V$
and the kinetic terms of the scalar fields which are given by
\begin{equation}\begin{aligned}\label{Ldef}
  e^{-1}\mathcal{L}
\ =\
\tfrac{1}{16}(D_{\mu}M_{MN})(D^\mu M^{MN})
+ \tfrac{1}{8}(D_{\mu}M_{\alpha\beta})(D^\mu M^{\alpha\beta})
- V(M,\xi,f)  + \ldots\ .
\end{aligned}\end{equation}
The covariant derivative of $M_{MN}$ reads
\begin{equation}
  \label{s2:DMMN}
  D_{\mu} M_{MN} = \partial_{\mu} M_{MN} +
2 {A_{\mu}}^{P\alpha} {\Theta_{\alpha P(M}}^Q M_{N)Q}\ ,
\end{equation}
where ${\Theta_{\alpha PM}}^Q= f_{\alpha MNP}-\xi_{\alpha[N}\eta_{P]M}$
is the matrix of gauge charges
and ${A_{\mu}}^{P+}$ are $n+6$ electric gauge bosons
while ${A_{\mu}}^{P-}$ are their magnetic duals.\footnote{In the embedding tensor
formalism electric and magnetic gauge bosons are simultaneously
introduced into the action
and a global $G=$SL(2)$\times$\SO(6,n) is manifest as long as $\xi^M_\alpha$ and $f_{\alpha MNP}$ transform as tensors under $G$. Any specific and consistent
choice of $\xi^M_\alpha$ and $f_{\alpha MNP}$ breaks that symmetry and determines the local gauge group
$G_0\subset G$.}
We see that a non-vanishing $\Theta_-$ leads  to magnetically charged
scalar fields but the above mentioned quadratic constraint
\eqref{eq:quadconstr} also
ensures mutual locality of electric and magnetic charges.
$D_\mu M_{\alpha\beta}$ depends only on $\xi_{\alpha M}$ which,
as we will see shortly, vanish for $N=4$ AdS backgrounds
implying that $\tau$ is uncharged and $D_\mu M_{\alpha\beta}$
reduces to an ordinary derivative.

The conditions for a supersymmetric AdS-background
can be concisely  formulated in terms of
the scalar components of the
$N=4$ supersymmetry transformations.
For the four gravitinos $\psi^i_\mu$,
the four spin-1/2
fermions in the gravitational multiplet $\chi^i$ and the gauginos $\lambda^i_a$
they are given by \cite{Schon:2006kz}
\begin{equation}\begin{aligned}\label{susytrans}
\delta \psi^i_\mu = &\ 2 D_\mu \epsilon^i - \tfrac23 A_1^{ij} \Gamma_\mu \epsilon_j + \dots \ , \\
\delta \chi^i = &\ - \tfrac43 \iu A^{ji}_2 \epsilon_j + \dots \ , \\
\delta \lambda^i_a = &\ 2 \iu A_{2aj}{}^i \epsilon^j + \dots \ ,
\end{aligned} \end{equation}
where $\epsilon_j$ are the four supersymmetry parameters and
the dots indicate terms that vanish in a maximally symmetric space-time
background.
The fermion shift matrices read
\begin{equation}\begin{aligned}\label{shiftM}
A^{ij}_1 \ = \ & \epsilon^{\alpha \beta} (\nu_\alpha)^* \nu_{kl}^M \nu_N^{ik} \nu_P^{jl} f_{\beta M}{}^{NP} \ , \\
A^{ij}_2 \ = \ & \epsilon^{\alpha \beta} \nu_\alpha \nu_{kl}^M \nu_N^{ik} \nu_P^{jl} f_{\beta M}{}^{NP} + \tfrac32\epsilon^{\alpha \beta} \nu_\alpha \nu_M^{ij} \xi^M_\beta \ , \\
A_{2ai}{}^j \ = \ & \epsilon^{\alpha \beta} \nu_\alpha \nu^M_a \nu^N_{ik} \nu_P^{jk} f_{\beta MN}{}^P - \tfrac14 \delta^j_i \epsilon^{\alpha \beta} \nu_\alpha \nu_a^M \xi_{\beta M} \ ,
\end{aligned} \end{equation}
where the $\nu_M^{ij}$ are defined with the help of SO(6) $\Gamma$-matrices
as
\be\label{nuGamma}
\nu_M^{ij}\ =\ \nu_M^m\,(\Gamma_m)^{ij}\ .
\ee
We give more details on the $\Gamma$-matrices in Appendix~\ref{N4}.
In terms of the shift matrices the scalar potential is given by
\begin{equation}
\label{s2:scalarpot}
 V= \tfrac 12 A_{2ai}{}^j A^*_{2aj}{}^i +\tfrac19 A_{2}{}^{ij} A^*_{2ij}
-\tfrac13 A_{1}{}^{ij}A^*_{1ij}{}\ .
\end{equation}

\section{Structure of $N=4$ AdS$_4$ backgrounds}\label{AdS}

In this section we study $N=4$ gauged supergravities
that admit a fully supersymmetric AdS$_4$ background, that is, all sixteen
supercharges are left
unbroken. The latter requirement demands that the supersymmetry variations
\eqref{susytrans}
of $\chi^i$ and $\lambda_a^i$ have to vanish in the AdS$_4$ background
while
the supersymmetry variations of the gravitinos  have to be
proportional to the cosmological constant.
Inspecting \eqref{susytrans} and \eqref{s2:scalarpot} we see that this implies
\be\label{susyshift}
\vev{A_2^{ij}}= \vev{A_{2a}^{ij}} =0 \ , \qquad \textrm{and} \qquad
\vev{A_1^{ij}  A^*_{1kj}} =  |\mu|^2\,\delta_k^i\ ,
\ee
where $\vev{V} = -\tfrac43|\mu|^2$ is the cosmological constant and $\vev{\cdot}$ indicates
that a quantity is evaluated in the AdS-background.
In  $A_2$  the first (second) term
is anti-symmetric (symmetric) in $i$ and $j$ and thus
they have to vanish independently.
Similarly in $A_{2a}$ the two terms correspond to a decomposition
into the trace and a traceless part and thus they also have to
vanish independently.
We can immediately conclude that fully supersymmetric
AdS$_4$ backgrounds can only occur in $N=4$ supergravities which have
\be\label{cxi}
 \xi_{\alpha}^M=0\ .
\ee
This property considerably simplifies the following analysis
and is also the reason why $M_{\alpha\beta}$ or similarly $\tau$ is uncharged in
the Lagrangian \eqref{Ldef}.
From \eqref{nutau}
we also see that for purely electric gaugings, i.e.\
$\xi_{\alpha}^M=f_{-NPQ}=0$, one has
$A_1=A_2$ and thus no supersymmetric AdS-background is possible \cite{deRoo:1986yw,Wagemans:1987zy}.

Inserting \eqref{cxi} into \eqref{shiftM}
the conditions \eqref{susyshift} simplify and read
\begin{eqnarray} 
\label{vgr}
\vev{A_1^{ij}} &  = & \vev{\nu_{kl}^M \nu_N^{ik} \nu_P^{jl}\nu_\alpha^*}\, \epsilon^{\alpha\beta}f_{\beta M}{}^{NP}
\  =\  \mu P^{ij} \ , \\
\vev{A_2^{ij}} &  = & \vev{\nu_{kl}^M \nu_N^{ik} \nu_P^{jl}\nu_\alpha} \,
\epsilon^{\alpha\beta}f_{\beta M}{}^{NP}
\   = \  0 \ , \label{vf}\\
\vev{A_{2a}^{ij}} &  = & \vev{\nu^M_a \nu^N_{ik} \nu_P^{jk}\nu_\alpha}\,
\epsilon^{\alpha\beta} f_{\beta MN}{}^{P}
\  =\    0 \ , \label{vga}
\end{eqnarray}
where $P^{ij}$ is a constant matrix obeying $P^{ik} P_{kj} = \delta^i_j$ but
is otherwise arbitrary.
The conditions \eqref{vgr}--\eqref{vga}
have to be solved subject to the quadratic
constraints \eqref{eq:quadconstr}  which for  $\xi^M_\alpha  = 0$
also simplify and are given by
\begin{equation}\label{eq:quadconstrN=4}\begin{aligned}
f_{\alpha R[MN}\, f_{\beta| PQ]}{}^R =  0 \ , \qquad
 \epsilon^{\alpha \beta} f_{\alpha MNR}\, f_{\beta PQ}{}^R =  0 \ .
\end{aligned} \end{equation}

Due to the homogeneity of the $N=4$ field space \eqref{N4coset}
one can translate any point of ${\cal M}$ to its origin and perform the analysis
there.\footnote{This has been frequently used, for example in  \cite{deRoo:2003rm,Borghese:2010ei,Horst:2012ub}.}
Here we prefer to perform the analysis at some arbitrary
but fixed vacuum expectation value
of the scalar fields corresponding to an AdS$_4$ background.
This leads us
to redefine the components of the embedding tensor
and introduce the complex quantities
\be\label{flcdef}
\f_{QRS} = 
\f_{1\,QRS}+\iu\f_{2\,QRS}=
\vev{\nu_{Q}^M \nu^N_R \nu^P_S\nu_\alpha}\, \epsilon^{\alpha\beta} f_{\beta NMP}\ ,
\ee
where  $\f_{1,2\,QRS }$ are the real and imaginary parts, respectively.
Using \eqref{nutau} they are given by
\be\label{f12}
 \f_{1\,QRS } =\vev{\tfrac{1}{\sqrt{\Im\tau}}\nu_{Q}^M \nu^N_R \nu^P_S}\,
\big(\vev{\Re\tau}\, f_{-NMP} -f_{+NMP}\big)\ ,\quad
  \f_{2\,QRS} =\vev{\sqrt{\Im\tau}\,\nu^M_Q \nu_{R}^N\nu^P_S}\,f_{-NMP}\ .
\ee
We see that $\f_2$ is directly related to the magnetic components $f_-$
of the embedding tensor
while $\f_1$ is an admixture of electric and magnetic components.
Note that at the origin of ${\cal M}$ the vielbeins are unit matrices,
$\vev{\Re\tau}$ vanishes and $\f_1$ is purely electric.
Before we proceed let us also give the quadratic constraint
\eqref{eq:quadconstrN=4} in terms of $\f$. Using \eqref{nutau} one finds
\begin{equation}\label{eq:quadconstr_fMNP}\begin{aligned}
 \f_{[MN}{}^R \f_{PQ]R} =  0 \ , \qquad
 \Re (\f_{[MN}{}^R \f_{PQ]R}^*) =  0 \ , \qquad
 \Im (\f_{MN}{}^R \f_{PQR}^*) =  0 \ .
\end{aligned} \end{equation}

Let us now turn to the solution of
the conditions \eqref{vgr}--\eqref{vga} and start by
analyzing the gaugino variation \eqref{vga}.
Using \eqref{nuGamma}, \eqref{flcdef} and the antisymmetry of the $f_{\alpha MNP}$
we can rewrite \eqref{vga} as
\begin{equation}
\f_{a mn}\, (\Gamma^{mn})^{ij} = 0 \ ,
\end{equation}
where
$\Gamma^{mn}$ are generators of $\SU(4)$ defined in Appendix~\ref{N4}.
Since the $\Gamma^{mn}$ are linearly independent generators we immediately
conclude
\begin{equation} \label{eq:condN=4I}
\f_{a mn}= 0 \ ,
\end{equation}
which, using \eqref{f12}, also implies $f_{\alpha amn}=0$.

We employ the same strategy to analyze the variations of the fermions
in the gravitational multiplet, i.e.\ \eqref{vgr} and \eqref{vf}.
Using  \eqref{nuGamma} and \eqref{flcdef} they are equivalent to
\be\label{vgrf2}
\f_{nmp}^*\,(\Gamma^n\Gamma^{*m}\Gamma^p)^{ij} =-\mu P^{ij}\ ,\qquad
\f_{nmp}\,(\Gamma^n\Gamma^{*m}\Gamma^p)^{ij} =0\ .
\ee
Since the antisymmetric products of three $\Gamma$-matrices are linearly
independent up to the relation \eqref{eq:rel3gamma},
we can further rewrite
\eqref{vgrf2} as
\begin{equation}\label{eq:condN=4II}
\f_{mnp} + \iu \epsilon_{mnpqrs} \f_{qrs} = 0 \ ,\qquad
\f_{mnp}\f^{*mnp} = 2|\mu|^2\ .
\end{equation}
We learn
that $\f_{mnp}$ ($\f^*_{mnp}$) is imaginary self-dual
(anti-self-dual) with a norm related to the
cosmological constant.\footnote{Note
that at the origin the $\f_{nmp}$ are related to the
$f^{(\pm)}$ defined in \cite{Borghese:2010ei}.
Furthermore, \eqref{eq:condN=4II} forbids
any real $\f_{mnp}$ which corresponds to the observation
that  a purely electric gauge theory does not admit an AdS-background.}
In addition to \eqref{eq:condN=4II} the $\f_{mnp}$
have to satisfy the quadratic constraints.
Due to \eqref{eq:condN=4I} the $mnpq$-component
of \eqref{eq:quadconstr_fMNP} simplifies and reads
\begin{equation}\label{eq:quadconstr_fmnp}\begin{aligned}
 \f_{[mn}{}^r \f_{pq]r} =  0 \ , \qquad
 \Re (\f_{[mn}{}^r \f_{pq]r}^*) =  0 \ , \qquad
 \Im (\f_{mn}{}^r \f_{pqr}^*) =  0 \ .
\end{aligned} \end{equation}
In Appendix~\ref{appB} we show that \eqref{eq:condN=4II} and
\eqref{eq:quadconstr_fmnp}  together
have a unique solution which can always be put
into the form
\begin{equation}\label{eq:structureconstantsc}
\f_{123} = \tfrac1{\sqrt 6}\,\mu \ , \qquad \f_{456} = -\tfrac\iu{\sqrt 6}\, \mu \ ,
\end{equation}
or, in terms of real and imaginary part and for  $\mu$ real\footnote{By an appropriate rotation of the gravitinos $\mu$ can always be chosen real.}
\begin{equation} \label{eq:structureconstants}
\f_{1\,123} =  \tfrac1{\sqrt 6}\, \mu \ , \qquad \f_{2\, 456} = -  \tfrac1{\sqrt 6}\,\mu \ .
\end{equation}
In terms of the $N=4$ gauge theory  this implies that
an $\textrm{SO}(3)_+\times \textrm{SO}(3)_-$   subgroup of the R-symmetry
group SO(6)$_R$ which rotates the six graviphotons into each other
has to be gauged in order for a fully supersymmetric AdS-background to exist.
One of the factors is electric while the second factor
is magnetic.

This concludes our solution of \eqref{vgr}--\eqref{vga} as they do not
involve the components $\f_{mab}$ and  $\f_{abc}$
of the (redefined) embedding tensor.
Or in other words we can choose
$\f_{ mab}=\f_{abc}=0$ without affecting the AdS solution.
In this case  the entire gauge group is
\be
G_0=\textrm{SO}(3)_+\times \textrm{SO}(3)_- \subset \textrm{SO}(6)_R\ ,
\ee
and the scalar fields are only charged with respect to the six graviphotons
while they remain neutral with respect to all other
$n$ Abelian vector fields. The number of vector multiplets $n$  in this
solution is arbitrary including $n=0$ in which case $\tau$
is the only scalar field.\footnote{For $n=0,1$ this solution was first found
in \cite{deRoo:1985jh} and it is also discussed in \cite{deRoo:2003rm}.}

However, as we will show now,
fully supersymmetric AdS-backgrounds with larger gauge groups $G_0$ can
also exist for $\f_{mab}\neq0$ and/or
$\f_{abc}\neq0$.\footnote{Physically the $f_{ \alpha mab}$
determine the supersymmetric fermionic and bosonic mass matrices
while  $f_{\alpha abc}$ only contributes to mass terms when supersymmetry is broken
\cite{Borghese:2010ei}.}
In this case the solution \eqref{eq:structureconstantsc}
of \eqref{vgr}--\eqref{vga} is unaffected
but the quadratic constraints \eqref{eq:quadconstr_fMNP}
change and have to be reanalyzed. In particular they couple different
components of the embedding tensor.

Let us first consider supergravities with $\f_{mab}=0,\f_{abc}\neq0 $.
In this case the quadratic constraints \eqref{eq:quadconstr_fMNP}
split into two disjoint set of conditions
and give a standard Jacobi-identity for
$\f_{abc}$. Thus the gauge group is
\be
G_0 = \textrm{SO}(3)_+ \times\textrm{SO}(3)_- \times G_0^v \subset \textrm{SO}(6,n)\ ,
\ee
where $G_0^v\subset \textrm{SO}(n)$
is the gauge group with structure constants  $f_{\alpha abc}$
which only acts among the gauge bosons of the vector multiplets.

For $\f_{mab}\neq0,\f_{abc}\neq0 $ the situation is slightly more involved.
First of all there can be a split within the
$\f_{abc}$ into two disjoint sets
so that one subset of them has no common indices
with $\f_{mab}$ and thus satisfies a standard Jacobi-identity
with no further interference terms in \eqref{eq:quadconstr_fMNP}.
As before this corresponds to a  separate factor
$G_0^v\subset \textrm{SO}(q), q\le n$
in the gauge group~$G_0$.

Now let us turn to the  $\f_{mab},\f_{abc}$ which do share common indices
$a,b,c$.
Considering the $mnab$-component of the last
constraint in  \eqref{eq:quadconstr_fMNP}
we learn that in the basis where
\eqref{eq:structureconstants} holds,
$\f_{1\,mab}$ can only be non-zero for $m=1,2,3$ while $\f_{2\,mab}$ can only
be non-zero for $m=4,5,6$. Furthermore the same equation also
says that $\f_{1\,mab}$ and $\f_{2\,mab}$ cannot share any $ab$ indices.
Or in other words the $\f_{mab}$ decompose
into two disjoint sets for $\f_{1\,mab}, m=1,2,3$ and $\f_{2\,nab},n=4,5,6 $.
With this observation all other quadratic constraints turn into
standard Jacobi-identities of three separate group factors
$G_+, G_-, G_0^v$
so that the total gauge group is of the form
\be
G_0 = G_+\times G_-\times G_0^v \subset \textrm{SO}(6,n)\ ,
\ee
where
\be
G_+ \subset  \textrm{SO}(3,m_+)\ ,\qquad
G_- \subset  \textrm{SO}(3,m_-)\ .\qquad
\ee
The maximal compact subgroups for each factor are
$\textrm{SO}(3)_\pm \times H_\pm$
with $ H_\pm\subset  \textrm{SO}(m_\pm)$ and $ m_++m_-+q=n$.
Special cases of this solution have been discussed in \cite{deRoo:2003rm}.
As we will see in the next section
in
the $N=4$ AdS background
 $G_0$ is spontaneously broken to its maximal compact subgroup.

\section{$N=4$ AdS moduli space}\label{moduli}

After having determined the $N=4$ AdS-backgrounds we turn to the question
to what extent they are isolated points in field space or if they can have
flat directions (a moduli space) which preserve all supercharges.
We use the same method as in \cite{deAlwis:2013jaa} in that we vary the
supersymmetry conditions \eqref{susyshift} and then find
all possible directions in the scalar field space~${\cal M}$
which are left undetermined by  \eqref{susyshift}.
More concretely, we look for continuous solutions of
\be
\delta A_1^{ij}= \delta A_2^{ij}= \delta A_{2a}^{ij} =0 \ ,
\ee
in the vicinity of a fully supersymmetric AdS$_4$ background.

In order to do so  we have to parametrize the variations of
the vielbeins. Let us define the $6n$ scalar field fluctuations
$\delta \phi_{ma}$ around the AdS$_4$ background value by
\begin{equation}
\delta \nu^m_M =  \langle\nu^a_M\rangle\, \delta \phi_{ma}\ .
\end{equation}
Then we find from \eqref{eq:vielbein_metric}
(suppressing henceforth the bracket $\langle \cdot\rangle$)
\begin{equation}\label{varnu}
\delta \nu^{a}_M = \nu^{m}_M\, \delta \phi_{ma} \ .
\end{equation}
Similarly, we have for the inverse vielbeins
\begin{equation}\label{varnuinv}
\delta \nu_{a}^M = - \nu^M_{m}\, \delta \phi_{ma}\ , \qquad
\delta \nu_m^M  = - \nu^M_a \,\delta \phi_{ma} \ .
\end{equation}
Thus at linear order in $\delta \phi$ the metric $M_{MN}$ is given by
\begin{equation} \label{eq:scalarmatrix}
 M_{MN} = \left( \begin{aligned}
 \delta_{mn} && 2 \delta \phi_{mb} \\ 2 \delta \phi_{an} && \delta_{ab}
 \end{aligned}
 \right) + {\cal O}(\delta \phi^2) \ .
\end{equation}
Similarly, for the $SL(2)/\SO(2)$ factor of ${\cal M}$ we have
\begin{equation}\label{vartau}
\delta \nu_\alpha = \tfrac{\iu}{2\Im \tau} ( \nu^*_\alpha \delta \tau - \nu_\alpha \delta \Re \tau )  \ .
\end{equation}
Using \eqref{varnuinv} and \eqref{vartau} we can also determine the
variations of $\f$ to be\footnote{These equations are  equivalent to the
gradient flow equations given in \cite{Bergshoeff:1985ms,D'Auria:2001kv}.}
\begin{equation} \label{eq:gradflow}\begin{aligned}
\delta \f_{npq} = & - 3 \delta_{m[n} \f_{pq]a}\, \delta\phi_{ma}
+ \tfrac{1}{\Im \tau}\, \Im \f_{npq}\, \delta\Re \tau - \tfrac{1}{2 \Im \tau}
\f^*_{npq}\delta\Im \tau \ ,\\[1ex]
\delta \f_{npb} = & (2 \delta_{m[n} \f_{p]ab} - \delta_{ab} \f_{mnp})\,  \delta\phi_{ma} + \tfrac{1}{\Im \tau}\, \Im \f_{npb}\, \delta\Re \tau - \tfrac{1}{2 \Im \tau}
\f^*_{npb}\,\delta\Im \tau \ .
\end{aligned} \end{equation}

With the help of these variations we can now discuss
the variations of $A_1$ and $A_2$. Starting from
\eqref{vgrf2} and using that $P^{ij}$ is constant  we obtain
\be
\delta \f^*_{mnp} = \delta \f_{mnp} = 0\ .
\ee
Using \eqref{eq:gradflow} and \eqref{eq:condN=4I}
this implies
\begin{equation}\label{taufixed}
 \mu P^{ij} \delta \tau = 0 \ ,
\end{equation}
leaving
 $\delta \tau = 0$ as the only solution. Thus the complex scalar $
\tau$ is necessarily fixed in any $N=4$ AdS-background.

We are left with the variation of $A_{2a}$ or in other words the variation of
\eqref{eq:condN=4I}. Using \eqref{eq:gradflow} and \eqref{taufixed} we find
\begin{equation} \label{eq:moduli1}
\f_{ mnp}\, \delta \phi_{pa} - 2 \f_{ab [m}\, \delta \phi_{n]b}  = 0 \ .
\end{equation}
In Appendix \ref{sec:so3} we show that all solutions of \eqref{eq:moduli1}
 have to be of the form
\begin{equation} \begin{aligned}\label{eq:moduli3}
 \delta \phi_{ma} &=  \f_{1 ab m} \lambda_1^{b} + \f_{2 ab m} \lambda_2^{b}
=  f_{\alpha ab m} \lambda_\alpha^b  \ ,
\end{aligned}\end{equation}
where $\lambda_{1,2}^{b}$ or equivalently  $\lambda^{ b}_\alpha$
are arbitrary real parameters.

The deformations \eqref{eq:moduli3} also have a geometrical meaning.
The Lagrangian and the background are invariant under global symmetry
transformations inside $SO(6,n)$ that leave the gauge group $G_0 =
G_+\times G_-\times G_0^v$ (and the associated structure constants)
invariant. This global symmetry is $G_0$ itself times its maximal commutant $H_c$ inside $SO(6,n)$.\footnote{Among the $Sl(2,\mathbb{R})$ transformations only $SO(2)$ preserves the electric and magnetic gaugings, which is a compact generator and therefore does not give any massless deformations.}
Since $H_c$ commutes with $G_\pm$, it commutes with $SO(3)_\pm$ and
therefore must be inside $SO(n)$, i.e.\ $H_c$ is a compact group.
Thus, the scalar deformations in \eqref{N4coset} which preserve
supersymmetry
should correspond to the non-compact directions in $G_0\times
H_c$. Since $G_0^v$ and $H_c$ are compact, the supersymmetric scalar
deformations span the coset
\begin{equation} \label{eq:coset}
{\cal M}_{N=4} = \frac{G_+}{H_+} \times \frac{G_-}{H_-} \ ,
\end{equation}
where $H_\pm$ are the compact subgroups of $G_\pm$ and contain an
$SO(3)_\pm$ factor. If we linearize the scalars in this coset, we
indeed find the deformations \eqref{eq:moduli3}.

Let us confirm the masslessness of the deformations \eqref{eq:moduli3}
by computing their scalar mass matrix.
The mass matrices of $N=4$ gauged supergravity have been given  and
analyzed for example in \cite{deRoo:2003rm,Borghese:2010ei}.
Nevertheless let us spend a few steps to derive them in an $N=4$ background
in our notation.
Inserting \eqref{shiftM} into
\eqref{s2:scalarpot} using \eqref{flcdef}  one finds for  $\xi_M=0$
\begin{equation}\label{eq:scalarpot2}
\tfrac14 V = -\tfrac13 \f_{mnp} ( \delta^{mq} \delta^{nr} \delta^{ps} + \iu \epsilon^{mnpqrs})\f^*_{qrs} + \tfrac12 \f_{amn} \f^*_{amn} + \tfrac19 \f_{mnp} ( \delta^{mq} \delta^{nr} \delta^{ps} - \iu \epsilon^{mnpqrs})\f^*_{qrs} \ .
\end{equation}
Computing the second derivative of
$V$ with respect to $\delta \phi_{am}$, we find for an $N=4$ background
(where \eqref{eq:condN=4I} and \eqref{eq:condN=4II} hold) the mass matrix
\begin{equation}\label{eq:massmatrix}\begin{aligned}
M_{am,bn} = & -16 \delta_{ab} \Re (\f_{mpq} \f^*_{npq} )+ 4 \Re[(2 \delta_{m[p} \f_{q]ac}  + \delta_{ac} \f_{mpq})(2 \delta_{n[p} \f^*_{q]bc}  + \delta_{bc} \f^*_{npq})] \\
=& - 16 \delta_{ab}  f_{\alpha mpq}  f_{\alpha npq}  + 4 (\delta_{ac} f_{\alpha mpq} + 2 f_{\alpha ca [p} \delta_{q]m})(\delta_{bc} f_{\alpha npq} + 2 f_{\alpha cb [p} \delta_{q]n}) \ .
\end{aligned}\end{equation}
We see that scalars of the form \eqref{eq:moduli3} indeed fulfill
$M_{am,bn}\delta \phi_{bn} = 0$ and thus are massless. This
confirmation can also be viewed as a consistency check of
\eqref{eq:moduli1}.\footnote{
Note that since \eqref{eq:massmatrix} is a sum of squares that come with different signs, there can be more massless scalars. Examples of this can be found for instance in \cite{deRoo:2003rm}.
However, such scalars do not preserve supersymmetry and will have a potential beyond quadratic order.}
It is easy to see from \eqref{eq:scalarpot2} that there are no mass terms mixing $\delta \tau$ and $\delta \phi_{am}$ for an $N=4$ vacuum.

Let us finally show that all flat directions in \eqref{eq:moduli3}
correspond to Goldstone bosons which are eaten by massive vector fields.
To do so we inspect the
covariant derivative of $\delta \phi_{ma}$ and
find from \eqref{s2:DMMN} and \eqref{eq:scalarmatrix}
\begin{equation}
D_\mu \delta \phi_{am} = \partial_\mu \delta \phi_{am} + \mu (A^{+ p}_\mu - A^{-p}_\mu) \epsilon_{pmn} \delta \phi_{an} + (A^{\alpha n}_\mu f_{\alpha nab} + A^{\alpha c}_\mu f_{\alpha cab}) \delta \phi_{bm} + 2 A^{\alpha b}_\mu f_{\alpha bam} + {\cal O}(\delta \phi^2) \ .
\end{equation}
First of all we note that in the AdS-background, i.e.\ for $\delta \phi_{am}=0$,
all six graviphotons $A^{\alpha m}$ are massless and thus, as expected,
the $\SO(3)_+\times \SO(3)_-$ part of the gauge symmetry is unbroken.
From the last term we see that there is a mass term for the gauge bosons
$A^{\alpha a}_\mu$ in the vector multiplets given by $\hat M^2_{\alpha a \beta b} \sim f_{\alpha acm}f_{\beta bcm}$.
This gives mass to ${\rm rk}(\hat M)$ gauge bosons where
$\hat M^{am}_{\alpha b} \sim f_{\alpha mab}$.
This
coincides with the number of flat directions determined in \eqref{eq:moduli3}
as the same matrix $\hat M$ appears.
This is no coincidence: When the gauge group is spontaneously broken
from $G_0 = G_+ \times G_- \times G_0^v \to H_+ \times H_- \times G_0^v$, the Goldstone bosons form the  coset \eqref{eq:coset}.
Therefore the supersymmetric directions in \eqref{eq:coset} are precisely the
Goldstone bosons
eaten by the massive  vectors.
Thus we showed that all supersymmetric  scalar deformations are Goldstone bosons
and therefore any AdS$_4$ background that preserves all supercharges
is an isolated point in field space with no further moduli.

\section*{Acknowledgments}

The work of J.L.\ was supported by the German Science Foundation (DFG) under
the Collaborative Research Center (SFB) 676 Particles, Strings and the Early
Universe, G.I.F.\ --
the German-Israeli Foundation for Scientific Research and Development
(GIF I-1-03847.7/2009) and the I-CORE program of the Planning and Budgeting
Committee and the Israel Science Foundation (grant number 1937/12).
He also thanks the Weizmann Institute and the Theory Group at CERN
for their kind hospitality during the course of  this work.

We have benefited from conversations and correspondence with Ofer Aharony,
Andrea Borghese, Murat G\"unaydin, Christoph Horst, Diederik Roest,
Henning Samtleben, Mario Trigiante and Marco Zagermann.

\newpage

\appendix
\noindent
{\bf\Large Appendix}
\section{$\SU(4)$ $\Gamma$-matrix properties }\label{N4}

The fermions of  $N=4$ gauged supergravity transform in the fundamental
representation of the $\SU(4)$ R-symmetry.  On the scalars
the R-symmetry acts
as $\SO(6) \sim \SU(4)/\mathbb{Z}_2$ rotations.
The two representations are linked via
the $\Gamma$-matrices $\Gamma_m^{ij}=\Gamma_m^{[ij]}$ given by \cite{Borghese:2010ei}
\begin{equation} \begin{aligned}
 \Gamma_1 = & \left( \begin{aligned}
 0 && 1 && 0 && 0\\
 -1 && 0 && 0 && 0 \\
 0 && 0 && 0 && 1 \\
 0 && 0 && -1 && 0
 \end{aligned} \right) \ ,
 \quad  \Gamma_2 = \left( \begin{aligned}
 0 && 0 && 1  && 0\\
 0 && 0 && 0 && -1 \\
 -1 && 0 && 0 && 0 \\
 0 && 1 && 0 && 0
 \end{aligned} \right) \ ,
  \quad \Gamma_3 = \left( \begin{aligned}
 0 && 0 && 0  && 1\\
 0 && 0 && 1 && 0 \\
 0 && -1 && 0 && 0 \\
 -1 && 0 && 0 && 0
 \end{aligned} \right) \ , \\[-2ex] \\
  \Gamma_4 = & \left( \begin{aligned}
 0 && \iu && 0 && 0\\
 -\iu && 0 && 0 && 0 \\
 0 && 0 && 0 && -\iu \\
 0 && 0 && \iu && 0
 \end{aligned} \right) \ ,
 \quad \Gamma_5 = \left( \begin{aligned}
 0 && 0 && \iu  && 0\\
 0 && 0 && 0 && \iu \\
 -\iu && 0 && 0 && 0 \\
 0 && -\iu && 0 && 0
 \end{aligned} \right) \ ,
  \quad   \Gamma_6 = \left( \begin{aligned}
 0 && 0 && 0  && \iu\\
 0 && 0 && -\iu && 0 \\
 0 && \iu && 0 && 0 \\
 -\iu && 0 && 0 && 0
 \end{aligned} \right) \ .
 \end{aligned} \end{equation}
They obey
\begin{equation} \label{eq:gammamatrices_cc}
 \{ \Gamma_m, \Gamma_n^*\} = 2 \delta_{mn} {\bf 1} \ ,
\qquad
(\Gamma_m)_{ij} = (\Gamma_m^{ij})^* = \tfrac12 \epsilon_{ijkl} \Gamma_m^{kl} \ .
\end{equation}
The antisymmetric products of two $\Gamma$-matrices $(\Gamma_{mn})^i_j = \tfrac12 (\Gamma_m)^{ik} (\Gamma^*_n)_{kj}$ are  the (linearly independent)
generators of $\SU(4)$. On the other hand the antisymmetric products of three $\Gamma$-matrices obey the relation
\begin{equation} \label{eq:rel3gamma}
 (\Gamma_{[m})^{ik} (\Gamma_n)_{kl} (\Gamma_{p]})^{jl} = \iu \epsilon_{mnpqrs} (\Gamma_q)^{ik} (\Gamma_r)_{kl} (\Gamma_s)^{jl}  \ .
\end{equation}
In \eqref{nuGamma} we use the $\Gamma$-matrices to convert the $\SO(6,n)$ vielbein components $\nu_M^m$ into objects with spinor indices, i.e.\ we define
$\nu_M^{ij} = \nu_M^m\Gamma_m^{ij}$.

\section{Classification of structure constants}\label{appB}
In this appendix we supply the details of the solution of
\eqref{eq:condN=4II} and \eqref{eq:quadconstr_fmnp}
given in \eqref{eq:structureconstantsc}.
We know from \eqref{eq:condN=4II} that $\f$ is the coefficient of an imaginary self-dual three-form, which means that the form is of type
$(2,1) \oplus (0,3)$ with respect to a given complex structure $I$ on the six-dimensional space parametrized by the index $m$. In addition we have
to solve the quadratic constraints \eqref{eq:quadconstr_fmnp} for such a three-form. If we write $\f$ with holomorphic and anti-holomorphic indices
$u,\bar u=1,2,3$, the quadratic constraints can be rewritten as
\begin{equation}\label{eq:quad_complex}\begin{aligned}
\f_{[\bar u \bar v |\bar x} \delta^{\bar x v}  \f_{u v |\bar w]} =   0 \ , \qquad
\f_{[u v |\bar v} \delta^{\bar v x}  \f_{w]x\bar u} =   0 \ ,  \qquad
\f_{\bar u \bar v \bar y}  \delta^{\bar y u} \f^*_{u \bar w \bar x} -  \f^*_{\bar u \bar v u}  \delta^{u \bar y} \f_{\bar w \bar x \bar y} =\ & 0\ , \\
 \f_{u v \bar w} \delta^{\bar w w} \f^*_{\bar u \bar v w} -  \f^*_{u v w} \delta^{w \bar w} \f_{\bar u \bar v \bar w} =  0 \ , \qquad
  \f_{w u \bar u } \delta^{w \bar w} \f^*_{\bar w \bar v v} -  \f^*_{u \bar u \bar w} \delta^{\bar w w} \f_{\bar v v w}   =\ &  0 \ ,\\
 2\f_{w [u [\bar u } \delta^{w \bar w} \f^*_{\bar v] v] \bar w} + 2\f^*_{\bar w[u [\bar u } \delta^{\bar w w} \f_{\bar v] v] w}
 + \f_{u v \bar w} \delta^{\bar w w} \f^*_{\bar u \bar v w} +  \f^*_{u v w} \delta^{w \bar w} \f_{\bar u \bar v \bar w}  =\ &  0 \ .
\end{aligned} \end{equation}
In terms of a more convenient parametrization
\begin{equation}\label{abdef}
 \f_{uv \bar u} = \epsilon_{uvw} \delta_{\bar u x} \alpha^{wx} \ , \qquad
 \f_{\bar u \bar v \bar w} =  \epsilon_{\bar u \bar v \bar w} \beta \ ,
\end{equation}
\eqref{eq:quad_complex} reads
\begin{equation}\label{eq:quad_complex2}\begin{aligned}
\beta \alpha^{[uv]} = 0 \ , \qquad \epsilon_{vwx} \alpha^{vu} \alpha^{wx} =  0 \ , \qquad
\alpha^{uv} \delta_{v \bar v} (\alpha^*)^{\bar u \bar v} = & | \beta|^2 \delta^{u \bar u} \ , \\
\alpha^{vu} \delta_{v \bar v} (\alpha^*)^{\bar v \bar u} + \epsilon^{uvw} (\alpha^*)_{vw} \epsilon^{\bar u \bar v \bar w}  \alpha_{\bar v \bar w} =& |\beta|^2  \delta^{u \bar u} \ , \\
\alpha^{uv} (\alpha^*)^{\bar v \bar u} - \alpha^{vu} (\alpha^*)^{\bar u \bar v} - \delta^{u \bar v} \delta_{w \bar w} \alpha^{wv} (\alpha^*)^{\bar w \bar u} + \delta^{\bar u v} \delta_{w \bar w} \alpha^{wu} (\alpha^*)^{\bar w \bar v}= & 0 \ .
\end{aligned} \end{equation}
As we will now show
these constraints imply that $\alpha$ is symmetric. To see this, assume that the antisymmetric part of $\alpha$ is non-zero and parametrize it by $\alpha^{[uv]} = \epsilon^{uvw} a_w$. From the last two equations of \eqref{eq:quad_complex2} we then find
\begin{equation}
a_u (\alpha^*)^{(\bar u \bar v)} = 2 \epsilon_{uvw} \delta^{v \bar w}  a^*_{\bar w} a_x \delta^{x(\bar u} \delta^{\bar v) w} \ ,
\end{equation}
which after contraction with $\bar a_{\bar u}$ and using the second equation of \eqref{eq:quad_complex2} shows that $a_u=0$ and $\alpha$ therefore is symmetric.
With this simplification the remaining conditions of \eqref{eq:quad_complex2}
imply
\begin{equation}\label{eq:quad_complex3}
\alpha^{uw}  (\alpha^*)_{wv}\  =\ | \beta|^2\, \delta^{u}_v \ ,
\end{equation}
where we lowered the indices with $\delta_{u \bar u}$. Since \eqref{eq:quad_complex3} says that the symmetric matrix $\alpha$ is also normal, we can diagonalize it as a complex matrix, and the diagonal entries are $\beta$ times some phase factors. Rotating the holomorphic coordinates by a phase corresponds to
$\SO(2)^3 \subset \SO(6)$ rotations, and together with electromagnetic $\SO(2)\subset \textrm{SL}(2)$ rotations
that only affect the overall phase of $\f$, we can bring $\alpha$ and $\beta$ into the final form
\begin{equation}
\beta = \tfrac1{4\sqrt 6}\, \mu \ , \qquad \alpha^{uv} = \tfrac1{4\sqrt 6}\, \mu \delta^{uv} \ ,
\end{equation}
where we also used the second equation in \eqref{eq:condN=4II}.
If we choose the complex structure as
\begin{equation}
I\ =\ \left( \begin{aligned}
0 && 0 && 0 && 0 && 0 && 1 \\
0 && 0 && 0 && 0 && 1 && 0 \\
0 && 0 && 0 && 1 && 0 && 0 \\
0 && 0 && -1 && 0 && 0 && 0 \\
0 && -1 && 0 && 0 && 0 && 0 \\
-1 && 0 && 0 && 0 && 0 && 0
\end{aligned} \right) \ ,
\end{equation}
we find for the embedding tensor  from \eqref{abdef}
\begin{equation}
\f_{123} =  \tfrac1{\sqrt 6}\,\mu \ , \qquad \f_{456} = -  \tfrac\iu{\sqrt 6}\, \mu \ .
\end{equation}
This indeed says that the gauge group of the graviphotons
is $\SO(3)_+ \times \SO(3)_- \subset \SO(6)$.

\section{$\SO(3)$ group theory}
\label{sec:so3}
In this appendix we determine the solution of
 \eqref{eq:moduli1}.
First let us record the $mnab$-component of the quadratic constraint
\eqref{eq:quadconstr_fMNP}
\begin{equation} \label{eq:quadconstrfmab}\begin{aligned}
&2 \f_{[ma}{}^c \f_{ bn]c}\ - \f_{mn}{}^p \f_{ abp} =0\ , \qquad
2 \Re(\f_{[ma}{}^c \f^*_{ bn]c}) 
- \Re(\f_{mn}{}^p \f^*_{ abp})
=0 , \\
& \Im (\f_{mn}{}^r \f_{abr}^*) =  0  \ , \qquad \Im (\f_{ma}{}^c \f_{nbc}^*)=  0 \ .
\end{aligned} \end{equation}
As we already noted in Section \ref{AdS},
the last but one equation together with
\eqref{eq:structureconstantsc} implies that $\f_{1mab}$
is only non-zero for $m=1,2,3$ while $\f_{2mab}$ is only non-zero for $m=4,5,6$.
This in turn simplifies  \eqref{eq:quadconstrfmab} further and
 implies the two decoupled equations
\begin{equation}\label{eq:fmab_commutator}
\f_{1,2\, mac} \f_{1,2\, ncb} - \f_{1,2\, nac} \f_{1,2\, mcb}
= \f_{1,2\, mn}{}^p \f_{1,2\, pab} \ ,
\end{equation}
for the real and imaginary part of $\f_{mab}$. From \eqref{eq:fmab_commutator}
we learn that the $\f_{1,2\, m ab}$ act on the $a$-type indices as
 $\SO(3)$ matrices
in some (possibly reducible) representations.

With these preliminaries let us return to the solution of \eqref{eq:moduli1}.
Since the real and imaginary parts of  \eqref{eq:moduli1} have to vanish independently, we can instead of \eqref{eq:moduli1} solve
\begin{equation}\label{eq:moduli1a}
\f_{1,2 \, mnp}\, \delta \phi_{pa} - 2 \f_{1,2 \, ab [m}\, \delta \phi_{n]b}  = 0 \ .
\end{equation}
Note again that due to \eqref{eq:structureconstantsc}
the equation \eqref{eq:moduli1a} for $\f_1$ gives a constraint for $\delta \phi_{ma}$ for $m=1,2,3$ while the equation with
$\f_2$ gives a constraint for $\delta\phi_{ma}$ with $m=4,5,6$.
We therefore  discuss the solution of \eqref{eq:moduli1a} only for $\f_1$
and then straightforwardly translate the result for $\f_2$.
In the following we thus omit the index~$1,2$.

Since the linear equation \eqref{eq:moduli1a}
is $SO(3)$ covariant it projects $\delta \phi_{ma}$ onto one or several $SO(3)$ representations. Let us pick a subspace of the $n$-dimensional vector space labeled by $a$ such that it forms one irreducible $m_i$-dimensional representation of $SO(3)$
(which we denote by $\rep{m_i}$), whose action is given by $\f_{mab}$ restricted to this subspace.
We label this subspace by the index $\tilde a=1,\dots, m_i$, so that $\f_{1m \tilde a \tilde b}$ acts on it transitively.
For this representation, \eqref{eq:moduli1a} reads
\begin{equation}\label{eq:moduli1subspace}
\f_{m\tilde a\tilde b} \delta \phi_{n \tilde b} - \f_{n\tilde a\tilde b} \delta \phi_{m \tilde b} = \f_{ mnp} \delta \phi_{p \tilde a} \ .
\end{equation}

Before we continue, let us recall some properties of
$\SO(3)$ representations. We denote the spin $s$ representation
with dimension $m=2s+1$ by ${\bf m}$. For the problem at hand we only need to consider
vector-like representations where $s$ is an integer.
The representation $\rep m$ can  then be understood as a totally
symmetric and traceless tensor of degree $s$.
They can be generated by tensor products of the form
\begin{equation} \label{eq:tensorprodSO3}
 {\bf 3} \otimes {\bf m} = {\bf m+2} \oplus {\bf m} \oplus {\bf m-2} \ , \qquad {\rm for}\ m \ge 2\ ,
\end{equation}
which are the symmetric traceless, anti-symmetric and trace components
of this tensor product, respectively.
A vector $v^{\tilde a}$ in the $\rep m$-representation can be written as $v_{n_1 \dots n_s} = l^{\tilde a}_{n_1 \dots n_s} v^{\tilde a}, n_1,\ldots,n_s=1,2,3$
where $l^{\tilde a}_{n_1 \dots n_s}$ is a constant symmetric, traceless tensor,
i.e.\ it obeys
$l^{\tilde a}_{n_1 \dots n_s} = l^{\tilde a}_{(n_1 \dots n_s)}$ and $l^{\tilde a}_{pq n_1 \dots n_{s-2}} \delta^{pq}=0$. The $\SO(3)$ action then reads
\begin{equation}\label{eq:SO3actionreps}
\f_{m \tilde a \tilde b} l^{\tilde b}_{n_1 \dots n_s} = - s \f_{mp (n_1}l^{\tilde a}_{n_2 \dots n_s) p} \ .
\end{equation}
This in particular means that $\SO(3)$ acts on the $\rep 3$-representation via the generators $-\f_{mnp}$.
From this definition of the $\f_{m \tilde a \tilde b}$ and \eqref{eq:structureconstantsc} we also find
\begin{equation}
 \f_{m \tilde a \tilde c} \f^*_{m \tilde c \tilde b} = \tfrac16\, s(s+1)\, |\mu|^2 \delta_{\tilde a\tilde b} \ .
\end{equation}
In this notation the representations \eqref{eq:tensorprodSO3} are then given by
\begin{equation}\label{eq:repsexplicit}\begin{aligned}
(w_m v^{\tilde a})_{s+1} =\ & l^{\tilde a}_{n_1 \dots n_s}  l^{\tilde b}_{(n_1 \dots n_s} w_{m)} v^{\tilde b} - \tfrac13 s\,  l^{\tilde a}_{m n_1 \dots n_{s-1}} l^{\tilde b}_{n_1\dots n_{s-1} p }w_p v^{\tilde b} \ , \\
(w_m v^{\tilde a})_{s} =\ & \tfrac{1}{s(s+1)}\, \f_{m n_1p}l^{\tilde a}_{n_2 \dots n_s p} \f^*_{qr (n_1 }  l^{\tilde b}_{n_2 \dots n_s) r}w_q v^{\tilde b} = \tfrac{1}{s(s+1)}\,  \f_{m\tilde a\tilde c} \f^*_{n\tilde c\tilde b} w_n v^{\tilde b} \ , \\
(w_m v^{\tilde a})_{s-1} =\ & \tfrac13 s\,  l^{\tilde a}_{m n_1\dots n_{s-1}} l^{\tilde b}_{n_1\dots n_{s-1} p }w_p v^{\tilde b}\ .
\end{aligned} \end{equation}

After this interlude let us return to the solution of \eqref{eq:moduli1subspace}.
If the representation is trivial, i.e.\ we have $m_i=1$ and $\f_{m \tilde a \tilde b}=0$, we find that \eqref{eq:moduli1subspace} implies $\delta\phi_{m \tilde a} =0$,
or in other words we do not find any moduli in this subspace.
 Let us therefore assume in the following that the representation ${\bf m_i}$ is non-trivial.
From their index structure we see that
the scalars $\delta\phi_{m \tilde a}$ are in the tensor product \eqref{eq:tensorprodSO3}.
We thus evaluate the condition \eqref{eq:moduli1a} for each representation in this tensor product individually.
If one uses \eqref{eq:repsexplicit} and \eqref{eq:fmab_commutator}, one can easily show that the anti-symmetric $\rep m_i$-representation obeys \eqref{eq:moduli1a}.
If $\delta \phi_{am}$ is in the ${\bf m_i+2}\oplus {\bf m_i-2}$ representation
we contract \eqref{eq:moduli1subspace} with $\f_{mnq}$ and find from \eqref{eq:structureconstantsc} and \eqref{eq:fmab_commutator} that
\begin{equation}
\Re (\f_{m\tilde a\tilde c} \f^*_{p\tilde c\tilde b}\,  -  \f_{p \tilde a \tilde c} \f^*_{m \tilde c\tilde b})\, \delta \phi_{p\tilde b}\ =\ \tfrac16 |\mu|^2\, \delta \phi_{m\tilde a} \ .
\end{equation}
From \eqref{eq:repsexplicit} we see that a $\delta \phi_{m \tilde a}$ in the ${\bf m_i+2}\oplus {\bf m_i-2}$ representation can be written as
\begin{equation}
\delta \phi_{m \tilde a} = l^{\tilde a}_{n_1 \dots n_s}  \delta \tilde \phi_{m n_1 \dots n_s} \ ,
\end{equation}
where $\tilde \phi$ is totally symmetric. Using \eqref{eq:SO3actionreps}
one can then show that $\f_{p\tilde c\tilde b} \delta\phi_{p\tilde b}=0$ and that
$\f_{p\tilde a\tilde c} \f_{m\tilde c\tilde b} \phi_{p\tilde b} = - \tfrac16 s |\mu|^2 \delta\phi_{m\tilde a}$. Thus, we can conclude that a $\delta\phi_{ma}$ in the
${\bf m_i+2}$ and ${\bf m_i-2}$ representations cannot fulfill \eqref{eq:moduli1subspace}.
Or in other words,
the condition \eqref{eq:moduli1subspace} projects onto the
$\rep m_i$-representation and we can parametrize
$\delta \phi_{m \tilde a} = \f_{m\tilde a\tilde b} \lambda^{\tilde b}$
with $\lambda^{\tilde b}$ being any element in that representation.

To summarize, we just showed that the scalar deformations
$\delta \phi_{m\tilde a}$ that preserve $N=4$ supersymmetry must be of the form
\begin{equation}
 \delta \phi_{ma} =  \f_{1 ab m} \lambda_1^{b} + \f_{2 ab m} \lambda_2^{b} \ ,
\end{equation}
corresponding to $\textrm{rk}(\hat M)= \sum_i m_{1\, i} + \sum_i m_{2\, i}$ massless degrees of freedom, where $\hat M^{am}_{\alpha b} \sim f_{\alpha mab}$.


\bibliography{GLSTV}
\providecommand{\href}[2]{#2}\begingroup\raggedright\endgroup

\end{document}